\documentclass[11pt]{article}
\usepackage{type1cm,amsmath,graphicx,indentfirst,amsthm}
\usepackage{amssymb}
\numberwithin{equation}{section}

\newcommand {\del}{\partial}

\newcommand{\tr}{\text{tr}}

\newcommand {\Hn}{\mathcal H_n}
\newcommand {\En}{\mathcal E_n}
\newcommand {\Fn}{\mathcal F_n}
\newcommand {\HH}{\mathcal H_4}
\newcommand {\EE}{\mathcal E_4}
\newcommand {\FF}{\mathcal F_4}
\newcommand {\WW}{\mathcal W_{4,3}}
\newcommand {\Xp}{\mathcal X^+}
\newcommand {\Xm}{\mathcal X^-}
\newcommand {\Xpm}{\mathcal X^\pm}
\newcommand{\RR}{\mathbb R}

\newcommand{\be}{\begin{equation}}
\newcommand{\bea}{\begin{eqnarray}}
\newcommand{\eea}{\end{eqnarray}}
\newcommand{\ba}{\begin{array}}
\newcommand{\ea}{\end{array}}
\newcommand{\ee}{\end{equation}}
\newcommand{\nn}{\nonumber}

\newtheorem*{conj*}{\textbf{Conjectures}}

\begin{document}
\begin{titlepage}

 \renewcommand{\thefootnote}{\fnsymbol{footnote}}
\begin{flushright}
 \begin{tabular}{l}
RUP-14-6, TUW--14--04
 \end{tabular}
\end{flushright}

 \vfill
 \begin{center}


\noindent{\Large \textbf{Unitary W-algebras and  three-dimensional higher \\[3mm] spin gravities with spin one symmetry}}\\
\vspace{1.5cm}

\noindent{ Hamid Afshar,$^a$\footnote{E-mail: h.r.afshar@rug.nl} Thomas Creutzig,$^{b}$\footnote{E-mail: creutzig@ualberta.ca} Daniel Grumiller,$^c$\footnote{E-mail: grumil@hep.itp.tuwien.ac.at} \\ Yasuaki Hikida$^d$\footnote{E-mail: hikida@rikkyo.ac.jp} and Peter B. R\o nne$^{e}$\footnote{E-mail: peter.roenne@uni.lu}}
\bigskip

 \vskip .6 truecm
\centerline{\it $^a$ Centre for Theoretical Physics, University of Groningen,} \centerline{\it Nijenborgh 4, 9747 AG Groningen, The
Netherlands}
\medskip
\centerline{\it $^b$ Department of Mathematical and Statistical Sciences, University of Alberta,} \centerline{\it Edmonton, Alberta T6G 2G1, Canada}
\medskip
\centerline{\it $^c$ Institute for Theoretical Physics, Vienna University of Technology,} \centerline{\it Wiedner Hauptstrasse 8-10/136, A-1040 Vienna, Austria}
\medskip
\centerline{\it $^d$Department of Physics, Rikkyo University,}
\centerline{\it  Toshima, Tokyo 223-8521, Japan}
\medskip
\centerline{\it $^e$ University of Luxembourg, Mathematics Research Unit, FSTC,
} \centerline{\it Campus
Kirchberg, 6, rue Coudenhove-Kalergi, }
\centerline{\it L-1359 Luxembourg-Kirchberg, Luxembourg}
 \vskip .4 truecm

 \end{center}

 \vfill
\vskip 0.5 truecm

\begin{abstract}
We investigate whether there are unitary families of W-algebras with spin one fields in the natural example of the Feigin--Semikhatov $W^{(2)}_n$-algebra.
This algebra is conjecturally a quantum Hamiltonian reduction corresponding to a non-principal nilpotent element.
We conjecture that this algebra admits a unitary real form for even $n$. Our main result is that this conjecture is consistent with the known part of the operator product algebra, and especially it is true for $n=2$ and $n=4$. Moreover, we find certain ranges of allowed levels where a positive definite inner product is possible.
We also find a unitary conformal field theory for every even $n$ at the special level $k+n=(n+1)/(n-1)$. At these points, the $W^{(2)}_n$-algebra is nothing but a compactified free boson.
This family of W-algebras admits an 't~Hooft limit that is similar to the original minimal model 't~Hooft limit.
Further, in the case of $n=4$, we reproduce the algebra from the higher spin gravity point of view.
In general, gravity computations allow us to reproduce some leading coefficients of the operator product.
\end{abstract}
\vfill
\vskip 0.5 truecm

\setcounter{footnote}{0}
\renewcommand{\thefootnote}{\arabic{footnote}}
\end{titlepage}

\newpage
\section{Introduction}

Higher spin gravity in three dimensions provides interesting toy models for quantum gravity, higher spin theories and aspects of holography, in the form of Anti-de~Sitter/conformal field theory (AdS/CFT) or more general gauge/gravity correspondences. Generalizing the seminal Brown--Henneaux analysis \cite{Brown:1986nw}, Henneaux and Rey \cite{Henneaux:2010xg}, and independently Campoleoni, Fredenhagen, Pfenninger and Theisen \cite{Campoleoni:2010zq}, showed that the asymptotic symmetry algebra for spin-3 gravity with AdS boundary conditions consists of two copies of $W_3$-algebras. Shortly afterwards, Gaberdiel and Gopakumar proposed a duality between $W_N$ minimal models in the large $N$ limit and families of 3-dimensional higher spin theories \cite{Gaberdiel:2010pz,Gaberdiel:2012uj}.

Most of the related early work remained focused on AdS holography, for instance in the discussion of higher spin black holes \cite{Gutperle:2011kf,Ammon:2012wc}, but it became soon clear that higher spin theories allow for more general holographic setups without introducing any additional matter fields \cite{Gary:2012ms}. Explicit constructions so far include Lobachevsky holography \cite{Afshar:2012nk}, Lifshitz holography \cite{Gutperle:2013oxa,Gary:2014} and flat space holography \cite{Afshar:2013vka,Gonzalez:2013oaa}. Typically, these more general holographic setups require to use non-principal embeddings of $sl(2)$ into $sl(n)$, and the ensuing asymptotic symmetry algebras are more complicated $W_n^{(m)}$ algebras, like the Polyakov--Bershadsky algebra \cite{Polyakov:1989dm,Bershadsky:1990bg} in the spin-3 Lobachevsky case \cite{Afshar:2012nk}. For large values of $n$ the number of non-principal embeddings grows exponentially with $n$, so that by sheer number these embeddings far outweigh the principal
one.

A universal property of all non-principal embeddings is the presence of at least one singlet in the wedge algebra, which translates into a current algebra as part of the asymptotic symmetry algebra. This current algebra has interesting implications for unitarity. Namely, Castro, Hijano and Lepage-Jutier argued \cite{Castro:2012bc} that at least semi-classically, i.e., in the limit of infinite central charge, the asymptotic symmetry algebra does not allow any unitary representations if there is a current algebra and a Virasoro algebra (as it is the case for all non-principal embeddings). The core argument was the observation that the level in the current algebra has opposite sign from the Virasoro central charge, so one of these two quantities necessarily has negative sign, thus implying the presence of negative norm states, at least for standard definitions of the vacuum and adjoint operators.

Exploiting properties of the Feigin--Semikhatov algebra $W_n^{(2)}$ \cite{Feigin:2004wb}, in \cite{Afshar:2012hc} it was shown that the no-go result of \cite{Castro:2012bc} can be circumvented by allowing the central charge to be arbitrarily large, but not infinite (the inequalities $0\leq c<n/4$ must hold, where $n$ can be arbitrarily large), and by additionally restricting the allowed values of the central charge to a specific discrete set that ensures non-negativity of the norm of all descendants of the vacuum.

For our purposes it is important to understand the source of the discreteness of the central charge, which is why we recall it now briefly. The Feigin--Semikhatov algebra $W_n^{(2)}$ is believed to be generated by  a current $J$, a Virasoro field $L$, higher spin fields $W^l$ [with $l=3..(n-1)$] and two additional fields $G^\pm$ that resemble a bosonic version of ${\cal N}=2$ supersymmetry generators. The commutator of the latter in general contains a central term that depends algebraically on the Virasoro central charge. Using the standard highest-weight definition of the vacuum and standard definitions of adjoint operators, it is then a simple exercise to show that generically half of the $G^\pm$-descendants of the vacuum have positive norm and the other half negative norm. The only exception arises when the aforementioned central term vanishes, which establishes a polynomial equation for the Virasoro central charge. This mechanism then leads to a discrete set of solutions for the Virasoro central charge
compatible with unitarity.

However, the construction reviewed above does not exclude the possibility to find a different vacuum or a different definition of adjoint operators where unitarity is less constrictive, i.e., does not restrict the central charge beyond convexity conditions. It is the main purpose of the present work to establish the existence of  such a scenario for gravity theories whose asymptotic symmetry algebras contain the Feigin--Semikhatov algebra $W_n^{(2)}$ for even $n$.

The main algebraic object of this work is the $W^{(2)}_n$-algebra introduced by Feigin and Semikhatov \cite{Feigin:2004wb}. They implicitly conjecture that this algebra is a quantum Hamiltonian
reduction of the affine vertex algebra of $sl(n)$ for a next to principal embedding of $sl(2)$ in $sl(n)$ (see e.g. \cite{MR2060475} for a general treatment of such reductions). The precise form of the conjecture has recently been formulated in \cite{Creutzig:2013pda}.
Support for this conjecture has been given in \cite{Creutzig:2011ah, Creutzig:2011an}, where the $W^{(2)}_n$-algebra at critical level has been constructed. At critical level the quantum Hamiltonian reduction is guaranteed to have a large center, and indeed also the $W^{(2)}_n$-algebra at critical level has such a large center. Our computations in the semi-classical limit actually provide further support for the correctness of this conjecture.

Let us recall what is known about the structure of the $W^{(2)}_n$-algebra. The algebra has
been defined by Feigin and Semikhatov as both a kernel of screening charges associated to simple roots of $sl(n|1)$
inside a free field theory and as a commutant or coset by the affine vertex algebra of $gl(n)$ of an extension of the affine vertex superalgebra of $gl(n|1)$.
The algebra is generated as a vertex algebra by two fields of conformal dimension $n/2$. These two fields together
with the dimension one field behave somehow similar as the affine vertex algebra of $sl(2)$. The superscript $(2)$ is due to this
resemblance. Moreover, the $W^{(2)}_2$ algebra is just the affine vertex algebra of $sl(2)$, and $W^{(2)}_3$ is the algebra
of Polyakov and Bershadsky \cite{Polyakov:1989dm,Bershadsky:1990bg}. Both are indeed the W-algebras corresponding to the next to principal embedding of $sl(2)$.
We ask the question, whether there exist exceptional levels for which the $W^{(2)}_n$-algebra defines a unitary algebra.
Recall, that the $W_n$-algebra, that is the quantum Hamiltonian reduction of affine $sl(n)$ for the principal embedding,
defines a unitary rational CFT if the level is exceptional in the sense that it takes values in a certain discrete set of rational
numbers. Kac and Wakimoto conjectured \cite{MR2452611}, that for every quantum Hamiltonian reduction there exist discrete sets of
allowed exceptional values of the level such that the corresponding W-algebra is a rational theory. This conjecture has been proven
by Tomoyuki Arakawa \cite{MR3096533} in the case of $W^{(2)}_3$.

Our idea is as follows. The second author gained some experience with the $W^{(2)}_n$-algebra and observed that indeed this
algebra behaves very much like affine $sl(2)$. For example, at critical level the classification of modules with finite-dimensional
zero-grade subspace was very analogous to the classification in the $sl(2)$ case \cite{Creutzig:2011ah}.
At a certain rational admissible level, the modules of the algebra behave as those of affine $sl(2)$ at fractional level \cite{Creutzig:2013pda}.
It is natural at least to us, to think about $W^{(2)}_n$ as a generalization of the WZW theory of $SL(2)$.
But recall, that the WZW models for positive integer level define unitary conformal field theories. These are of course
based on the unitary real forms, for example the unitary form $SU(2)$ of $SL(2,\mathbb C)$. Our idea is thus to try to find
a unitary real form of $W^{(2)}_n$ proceeding as much as possible in analogy to $su(2)$. Indeed, our main technical result is that
such a form exists, at least as far as the operator product algebra is known.

This work is organized as follows.
In section \ref{se:2} we state our main results, that summarize the outcome of the technical computations of the following sections.
In section \ref{sec:FS} we review salient features of the Feigin--Semikhatov algebra $W_n^{(2)}$.
In section \ref{sec:FSunitary} we provide a unitary real form that differs from the one previously used and discuss implications for unitarity.
In section \ref{sec:c1} we focus on a special value of the 't~Hooft coupling that lies at the boundary of the interval permitted by the unitarity analysis of the previous section.
We find that at that value the simple $W^{(2)}_n$ conformal field theory is nothing but a unitary lattice CFT.
In section \ref{sec:gravity} we discuss the higher spin gravity perspective.
We then terminate with a short outlook.

\section{Results}\label{se:2}

Our main technical result is that the $W^{(2)}_n$-algebra of level $k$ of Feigin and Semikhatov seems to allow for a unitary real form if $n$ is even.
We can verify this statement only as far as the operator product algebra of the Feigin--Semikhatov algebra is known.
In section \ref{sec:FS} we recall the known operator product algebra. Especially we can provide the complete algebra in the case $n=4$.
It turns out that our conjecture is compatible with the operator product algebra, and especially it is true for $n=2$ and $n=4$.
We provide the necessary computations for this in section \ref{sec:FSunitary}.
There is also one special level $k+n=(n+1)/(n-1)$, where we even find a unitary conformal field theory for all even $n$ with symmetry algebra looking like
a simple quotient of $W^{(2)}_n$, see section \ref{sec:c1}. This means, the symmetry algebra is a simple algebra, and its operator product algebra agrees with the known algebra of
$W^{(2)}_n$ at that level modulo a vertex algebra ideal.
Note that for odd $n$, there are bosonic fields of half integral spin, so the spin statistic relations do not hold. In these cases it seems to be impossible to find a positive definite inner product (except for the discrete set of levels discussed in \cite{Afshar:2012hc}).

Having this result, the next question is for which level $k$ does this algebra admit a positive definite inner product.
We can check this question on the level of the inner product of the generating fields of the W-algebra.
Details of the computation are again outlined in section \ref{sec:FSunitary}.
The case $n=2$ is just the affine vertex algebra of $su(2)$ and it is well-known that a positive inner product can only exist for $k\geq 0$. Note, that in this case
it is also well-known that one even obtains a unitary conformal field theory if $k$ is a positive integer.
In the case $n=4$, we find that a positive definite inner product can exist if $k$ takes values in a certain interval, namely
\begin{equation}
\frac{4}{3} \leq k+4 \leq \frac{15}{8}\,.
\end{equation}
For general even $n$, the known operator product algebra restricts the possible values of the level
for a positive inner product to satisfy
\begin{equation}
 \frac{n}{n-1}\leq k+n\leq \frac{n^2-1}{n(n-2)}.
 \end{equation}
We are interested in an 't~Hooft limit. For this we define a coupling constant $\lambda$ by
\begin{equation}\label{eq:lambda}
\lambda=(n-1)(k+n-1).
\end{equation}
Then the condition for a unitary W-algebra translates to
\begin{equation}
 1\leq \lambda\leq \frac{(2n-1)(n-1)}{n(n-2)}=2+\frac{(n+1)}{n(n-2)}\,.
\end{equation}
For a higher spin duality, we require the central charge to scale with $n$. Indeed, we find that
\begin{equation}
\begin{split}
 c&=(1-\lambda)\frac{1-2n+\lambda n \frac{(n-2)}{(n-1)}}{1+\lambda/(n-1)} \sim (\lambda-1)(2-\lambda)n\,.
\end{split}
 \end{equation}
In the large $n$ limit, the allowed range for $\lambda$ is the interval $[1, 2]$ and hence the coefficient $(\lambda-1)(2-\lambda)$ is positive.

Let us comment on the level for the original higher spin W-algebra correspondence.
There $\lambda=n/(\tilde k+n)$ is related to the level $\tilde k$ of the coset construction. The relation to the level $k$
of the affine Lie algebra  whose quantum Hamiltonian reduction it is, is
$$
k+n=\frac{n}{\lambda +n}\leq 1\,.
$$
Especially we see that only for small levels we get a unitary rational CFT.
Further, the allowed $\lambda$ that lead to a unitary CFT form a discrete subset $\mathbb Q$ of the interval $[0, 1]$.
This situation is similar to what we have
above for $W^{(2)}_n$.

In section \ref{sec:gravity}, we then first compute the algebra of $sl(4, \RR)$ Chern--Simons theory with next to principal boundary conditions. We find nice agreement with the unitary $W^{(2)}_4$-algebra. We are also able to compute some operator product coefficients for general $n$, and again find nice agreement. These computations both support the conjecture that the $W^{(2)}_n$-algebra of Feigin and Semikhatov is a quantum Hamiltonian reduction as well that it is the holographic dual of a higher spin algebra with spin one symmetry.

Let us summarize in formulating our main conjectures.
\begin{conj*} ${}$ \\[-5mm]
 \begin{enumerate}
\item For every $\lambda\in [1, 2]$ there is a unitary simple real form of $W^{(2)}_{2n}$ at level $k$ with positive inner product.
\item The semi-classical limit of the $W^{(2)}_{2n}$-algebra is realized by $sl(2n, \RR)$ Chern--Simons theory with next to principal boundary conditions.
\item The unitary $W^{(2)}_{2n}$-algebra is the holographic dual of a higher spin algebra with spin one symmetry.
\end{enumerate}
\end{conj*}

\section{The $W^{(2)}_n$ algebra of Feigin and Semikhatov}\label{sec:FS}

We start our considerations with what is known about the Feigin--Semikhatov $W^{(2)}_n$ algebra of level $k$.
This algebra is constructed in \cite{Feigin:2004wb} as both a coset of supergroup type and as the chiral algebra of
a conformal field theory associated to a set of screening charges of supergroup type inside a lattice super algebra.
Both constructions allow, in principle, for a computation of the operator algebra, though this is very tedious and Feigin and Semikhatov provide the leading
contributions of important operator products.
Here, we recall this data.
The central charge is
\begin{equation}
c_n(k) =-\frac{\left((k+n)(n-1)-n\right)\left((k+n)(n-2)n-n^2+1\right)}{(k+n)}\,.
\label{eq:c}
\end{equation}
The algebra contains a Virasoro field $T(z)$ with this central charge. As a vertex algebra it is generated by two dimension $n/2$ fields, $\mathcal E_n$ and $\mathcal F_n$.
It also contains one $u(1)$-current $\mathcal H_n$.
We list now the known operator products for general $n$
\begin{equation}
\begin{split}
\Hn(z)\Hn(w) &\sim \frac{\ell_n(k)}{(z-w)^2}\,, \quad
\Hn(z)\En(w) \sim \frac{\En(w)}{(z-w)}\,,\quad   \Hn(z)\Fn(w) \sim -\frac{\Fn(w)}{(z-w)}
\end{split}
\end{equation}
and
\begin{equation}
\begin{split}
\En(z)\Fn(w) &\sim \frac{\lambda_{n-1}(n,k)}{(z-w)^n} +\frac{n\lambda_{n-2}(n, k)\Hn(w)}{(z-w)^{n-1}} +\frac{\lambda_{n-3}(n, k)}{(z-w)^{n-2}} A(w)\\
&\quad +\frac{\lambda_{n-3}(n, k)}{(z-w)^{n-3}} B(w)+\frac{\lambda_{n-2}(n, k)}{(z-w)^{n-3}} C(w)+\dots
\end{split}
\label{eq:angelinajolie}
\end{equation}
with coefficients
\begin{equation}
\begin{split}
\ell_n(k)&= \frac{(n-1)(k+n)-n}{n}\,,\qquad
\lambda_m(n, k) =\prod_{i=1}^m \left(i\left(k+n-1\right)-1\right)
\end{split}
\end{equation}
and normal ordered products
\begin{equation}\nonumber
\begin{split}
A(w) &= \frac{n(n-1)}{2} \Hn(w)\Hn(w)+\frac{n\left(\left(n-2\right)\left(k+n-1\right)-1\right)}{2} \partial \Hn(w) -(k+n) T(w) \\
B(w) &= \mathcal W_{n, 3}(w) -(k+n)\left(\frac{1}{2} \partial T_\perp(w) +\frac{1}{\ell_n(k)}\Hn(w)T_\perp(w)\right) \\
C(w) &= \frac{n}{6\ell_n(k)^2} \Hn(w)\Hn(w)\Hn(w) +\frac{n}{2\ell_n(k)}\partial \Hn(w) \Hn(w) +\frac{n}{6} \partial^2\Hn(w)\\
T_\perp(w) &= T_n(w) - \frac{1}{2\ell_n(k)} \Hn(w)\Hn(w)\,.
\end{split}
\end{equation}
Here $\mathcal W_{n, 3}$ is a dimension three primary field.
In the analysis of \cite{Afshar:2012hc} the quantity $\lambda_{n-1}(n,k)$ was required to vanish, which eliminates the anomalous term in the OPE of $\En(z)$ and $\Fn(w)$ \eqref{eq:angelinajolie}. In the present work we shall not find it necessary to impose such a restriction.

The $W^{(2)}_4$ algebra will be our main example. In this case, complete operator products are known.
We start with the one of $\EE$ with $\FF$,
\begin{equation}
\begin{split}
\frac{1}{(k+2)}&\EE(z)\FF(w) \sim \frac{(2k+5)(3k+8)}{(z-w)^4}+\frac{4(2k+5)\HH(w)}{(z-w)^3}\\
& +  \frac{-(k+4)T(w)+6\HH(w)\HH(w)+2(2k+5)\partial\HH(w)}{(z-w)^2}+\frac{1}{(z-w)}\Big( \WW(w) \\
&\left.-\frac{k+4}{2} \partial T(w) -\frac{4(k+4)}{3k+8}T(w)\HH(w)+
\frac{8(11k+32)}{3(3k+8)^2}\HH(w)\HH(w)\HH(w) \right. \\
&+6\partial \HH(w)\HH(w)+\frac{4(26+17k +3k^2)}{3(3k+8)}\partial^2\HH(w)\Big)\,.
\end{split}
\end{equation}
Using the short-hand notation $\EE=\Xp$ and $\FF=\Xm$, the OPE of $\WW$ with $\Xpm$ can be compactly written as
\begin{equation}
\begin{split}
\WW(z)\Xpm&(w) \sim \pm \frac{2(k+4)(3k+7)(5k+16)}{(3k+8)^2}\frac{\Xpm(w)}{(z-w)^3}+\frac{1}{(z-w)^2}\times\\
&\times\left( \pm 3 \frac{(k+4)(5k+16)}{2(3k+8)}\partial\Xpm(w) - 6 \frac{(k+4)(5k+16)}{(3k+8)^2}\HH(w)\Xpm(w)\right) \\
& -\frac{k+4}{k+2}\frac{1}{(z-w)}\left(\frac{8(k+3)}{3k+8}\HH(w)\partial\Xpm(w) + \frac{4(3k^2+15k+16)}{(3k+8)^2}\times\right.\\
&\left.\times\partial\HH(w)\Xpm(w)\mp
(k+3)\partial^2\Xpm(w) \pm \frac{2(k+4)}{3k+8}T(w)\Xpm(w) \right. \\
&\left.\mp \frac{4(5k+16)}{(3k+8)^2}\HH(w)\HH(w)\Xpm(w) \right)\,.
\end{split}
\end{equation}
Finally, the dimension three field has the following OPE with itself
\begin{equation}
\begin{split}
\WW(z)\WW(w)&\sim \frac{2(k+4)(2k+5)(3k+7)(5k+16)}{3k+8}\frac{1}{(z-w)^6}\\
&-\frac{(k+4)^2(5k+16)}{3k+8}\frac{3T_\perp(w)}{(z-w)^4}-\frac{(k+4)^2(5k+16)}{2(3k+8)}\frac{3\partial T_\perp(w)}{(z-w)^3}\\
&+\frac{1}{(z-w)^2}\left(-\frac{3(k+4)^2(5k+16)(12k^2+59k+74)}{4(3k+8)(20k^2+93k+102)}\partial^2 T_\perp(w)  \right. \\
&\left. +\frac{8(k+4)^3(5k+16)}{(3k+8)(20k^2+93k+102)}T_\perp(w)T_\perp(w)+4(k+4)\Lambda(w) \right)\\
&\frac{1}{(z-w)}\left( -\frac{(k+4)^2(5k+16)(12k^2+59k+74)}{6(3k+8)(20k^2+93k+102)}\partial^3 T_\perp(w)  \right. \\
&\left. +\frac{8(k+4)^3(5k+16)}{(3k+8)(20k^2+93k+102)}\partial T_\perp(w)T_\perp(w)+2(k+4)\partial\Lambda(w)\right),
\end{split}
\end{equation}
where
\begin{equation}
T_\perp(w) = T(w)-\frac{2}{3k+8}\HH(w)\HH(w)
\end{equation}
and $\Lambda(w)$ is a dimension four Virasoro primary field, but a W-algebra descendant. It is the following normally ordered product in the strong generators of the W-algebra
\begin{align}\nonumber
&(k+2)^2\Lambda(w) = \Xp(w)\Xm(w) -\frac{k+2}{2}\partial \WW(w) -\frac{4(k+2)}{3k+8}\WW(w)\HH(w) \\ \nonumber
&+ \frac{3(k+2)^2(k+4)(6k^2+33k+46)}{2(3k+8)(20k^2+93k+102)}\partial^2 T_\perp(w)  
-\frac{(k+2)(k+4)^2(11k+26)}{2(3k+8)(20k^2+93k+102)}T_\perp(w)T_\perp(w)\\ 
&+ \frac{2(k+2)(k+4)}{(3k+8)}\partial\left(T_\perp(w)\HH(w)\right) +
\frac{8(k+2)(k+4)}{(3k+8)^2}T_\perp(w)\HH(w)\HH(w)\\ \nonumber
&- \frac{8(k+2)(2k+5)}{3(3k+8)}\partial^2\HH(w)\HH(w) -\frac{2(k+2)(2k+5)}{(3k+8)}\partial\HH(w)\partial \HH(w)\\ \nonumber
&- \frac{16(k+2)(2k+5)}{(3k+8)^2}\partial \HH(w)\HH(w)\HH(w)\\ \nonumber
&-\frac{32(k+2)(2k+5)}{3(3k+8)^3}\HH(w)\HH(w)\HH(w)\HH(w)-\frac{(k+2)(2k+5)}{6} \partial^3\HH(w).
\end{align}

\section{A unitary real form of $W^{(2)}_{n}$ for $n$ even}\label{sec:FSunitary}

In this section, we argue that the Feigin--Semikhatov algebra admits a unitary real form in the case of even  $n$.
As the complete operator product algebra is only known for $n=1,2,3,4$, we can only prove this statement in the cases $n=2,4$.
But we strongly believe that it holds in general. We start by a case by case analysis.

\subsection{The unitary $W^{(2)}_2$-algebra}\label{sec:sl2}

The case of $n=2$ is nothing but the affine vertex algebra of $sl(2)$.
The natural real form of $sl(2)$ has basis $e, f$ and $h$ with
\begin{equation}
[e,f]=2h\,, \qquad [h,e]=e\quad  \text{and}\quad [h,f]=-f\,.
\end{equation}
The corresponding vertex algebra has generating fields $e(z), f(z)$ and $h(z)$ with operator products
\begin{equation}
\begin{split}
 h(z)h(w)&\sim \frac{k/2}{(z-w)^2}\,,\quad\ \ e(z)f(w)\sim  \frac{k}{(z-w)^2}+\frac{2h(w)}{(z-w)}\\
h(z)e(w)&\sim \frac{e(w)}{(z-w)}\,,\qquad h(z)f(w)\sim -\frac{f(w)}{(z-w)}
\end{split}
\end{equation}
and $k$ is the level. The central charge is $c=3k/(k+2)$. The argument of \cite{Afshar:2012nk,Afshar:2012hc} for
unitarity is that there are no states of negative norm, requiring $k=0$ and hence also $c=0$.
But there is a way out, namely consider the unitary real form $su(2)$ with generators
$X=i(e+f)/2, Y=(e-f)/2$ and $J=ih$. Then the commutation relations are
\begin{equation}
 [X,Y]=-J\,, \qquad [J,X]=-Y\,,\qquad [J,Y]=X\,.
\end{equation}
Operator products of corresponding currents are
\begin{equation}
\begin{split}
X(z)X(w)&\sim Y(z)Y(w)\sim J(z)J(w)  \sim \frac{-k/2}{(z-w)^2}\,,\\
\qquad X(z)Y(w)&\sim -\frac{J(w)}{(z-w)}\,, \quad
J(z)X(w)\sim -\frac{Y(w)}{(z-w)}\,,\quad
J(z)Y(w)\sim \frac{X(w)}{(z-w)}\,.
\end{split}
\end{equation}

The natural bilinear form is the Killing form and this is negative definite
hence for the corresponding modes of currents we need to define the adjoint as
\begin{equation}
X^\dagger_m=-X_{-m}\,,\qquad  Y^\dagger_m=-Y_{-m}\,,\qquad  J^\dagger_m=-J_{-m}\, ,
\end{equation}
which directly follows from the $\mathbb{Z}_2$ anti-automorphism $h_m^\dagger=h_{-m}$, $e_m^\dagger=f_{-m}$.
Thus the unitarity requirement is that $k\geq 0$. Indeed for positive integer $k$ it is known that the underlying conformal field theory, the WZW model, is a unitary rational theory.
The point of recalling this well-known fact is that it motivates a generalization.

\subsection{The unitary $W^{(2)}_4$-algebra}\label{sec:w4}

We will now show, that the $W^{(2)}_4$-algebra admits a real form, that resembles very much the unitary real form of affine $su(2)$,
hence we will call this real form the unitary $W^{(2)}_4$-algebra.
Define in analogy to last section
\begin{equation}\label{eq:scarlett}
X_n=\frac{i(\mathcal E_n+\mathcal F_n)}{2\sqrt{\lambda_{n-2}(n, k)}}, \qquad Y_n=\frac{(\mathcal E_n-\mathcal F_n)}{2\sqrt{\lambda_{n-2}(n, k)}},\qquad  J_n=i\mathcal H_n\, ,
\end{equation}
and also $Z_n=i\mathcal W_{n, 3}$.
The normalization means that operator product expansion coefficients become rational functions in the level with possible poles at the zeros of $\lambda_{n-2}(n, k)$. Hence for these values of the level one sould use a different normalization. However, the cases where $\lambda_{n-2}(n, k)$ vanishes are exactly the discrete levels investigated in \cite{Afshar:2012hc}. They lead to unitary real forms of the $W^{(2)}_n$ algebra provided that the central charge is non-negative. In general poles and zeroes of operator product coefficients indicate that the universal W-algebra might not be simple and in such a situation one should study its simple quotient.

We use the same normal ordering convention and short-hand notation as Feigin and Semikhatov in \cite{Feigin:2004wb}, that is for three fields $A(w), B(w), C(w)$ we have
\begin{equation}
A(w)B(w) \ \stackrel {\text{def}}{ =} \  :A(w)B(w):\quad\text{and}\quad A(w)B(w)C(w)  \ \stackrel {\text{def}}{ =} \ :A(w) (:B(w)C(w):):\,.
\end{equation}
Then the non-regular opertator products involving the $u(1)$-current $J_4$ are
\begin{equation}
 \begin{split}
J_4(z)J_4(w) &\sim  -\frac{1}{4}\frac{(3k+8)}{(z-w)^2}\,,\quad
J_4(z)X_4(w) \sim -\frac{Y_4(w)}{(z-w)}\,,\quad
J_4(z)Y_4(w) \sim \frac{X_4(w)}{(z-w)}\,.
\end{split}
\end{equation}
Those of the fields $X_4, Y_4$ with each other are
\begin{align}\nonumber
X_4(z)X_4(w) &\sim Y_4(z)Y_4(w) \sim -\frac{(3k+8)}{2(z-w)^4}+\frac{\bigl((k+4)T(w)+6J_4(w)J_4(w)\bigr)}{2(2k+5)(z-w)^2} \\
&\qquad\qquad\quad +\frac{\bigl((k+4)\partial T(w)+12\partial J_4(w) J_4(w)\bigr)}{4(2k+5)(z-w)},\\ \nonumber
X_4(z)Y_4(w) &\sim -\frac{2J_4(w)}{(z-w)^3}-\frac{\partial J_4(w)}{(z-w)^2}-
\frac{1}{2(2k+5)(z-w)}\left(Z_4(w)-\frac{4(k+4)T(w)J_4(w)}{(3k+8)}\right.\\ \nonumber
&\left.-\frac{8(11k+32)}{3(3k+8)^2}J_4(w)J_4(w)J_4(w)
+\frac{4(26+17k+3k^2)}{3(3k+8)}\partial^2J_4(w)\right).
\end{align}
The operator product of $X_4$ with the dimension three primary $Z_4$ is
\begin{equation}\nonumber
 \begin{split}
Z_4(z)&X_4(w) \sim -\frac{2(k+4)(3k+7)(5k+16)}{(3k+8)^2}\frac{Y_4(w)}{(z-w)^3}-
\frac{3(k+4)(5k+16)}{2(3k+8)^2}\\
&\quad \times\frac{(3k+8)\partial Y_4(w)-4 J_4(w) X_4(w)}{(z-w)^2}-
\frac{(k+4)}{(k+2)(3k+8)^2}\frac{1}{(z-w)}\Bigl( 8(k+3)\\
&\quad\times(3k+8)J_4(w)\partial X_4(w)
+4(3k^2+15k+16)\partial J_4(w)X_4(w)+(k+3)(3k+8)^2 \partial^2 Y_4(w)\\
&\quad -2(k+4)(3k+8) T(w)Y_4(w)- 4(5k+16)J_4(w)J_4(w)Y_4(w)\Bigr)
\end{split}
\end{equation}
and the one of $Y_4$ with $Z_4$ is
\begin{equation}\nonumber
 \begin{split}
Z_4(z)&Y_4(w) \sim \frac{2(k+4)(3k+7)(5k+16)}{(3k+8)^2}\frac{X_4(w)}{(z-w)^3}+
\frac{3(k+4)(5k+16)}{2(3k+8)^2}\\
&\quad \times\frac{(3k+8)\partial X_4(w)+4 J_4(w) Y_4(w)}{(z-w)^2}-
\frac{(k+4)}{(k+2)(3k+8)^2}\frac{1}{(z-w)}\Bigl(8(k+3) \\
&\quad \times(3k+8)J_4(w)\partial Y_4(w)
+4(3k^2+15k+16)\partial J_4(w)Y_4(w)-(k+3)(3k+8)^2 \partial^2 X_4(w)\\
&\quad +2(k+4)(3k+8) T(w)X_4(w)+ 4(5k+16)J_4(w)J_4(w)X_4(w)\Bigr).
\end{split}
\end{equation}
Finally the operator product of $Z_4$ with itself is
\begin{equation}\nonumber
 \begin{split}
Z_4(z)&Z_4(w) \sim -\frac{2(k+4)(2k+5)(3k+7)(5k+16)}{(3k+8)(z-w)^6}+\frac{(k+4)^2(5k+16)}{(3k+8)}\frac{3 T_\perp (w)}{(z-w)^4}\\
&\quad+ \frac{(k+4)^2(5k+16)}{2(3k+8)}\frac{3 \partial T_\perp (w)}{(z-w)^3}+\frac{1}{(z-w)^2}\Bigl( \frac{3(k+4)^2(5k+16)(12k^2+59k+74)}{4(3k+8)(20k^2+93k+102)}\partial^2 T_\perp(w)\\
&\quad-\frac{8(k+4)^3(5k+16)}{(3k+8)(20k^2+93k+102)}T_\perp(w)T_\perp(w)-4(k+4)\Lambda(w)\Bigr)
+\frac{(k+4)}{(z-w)}\Bigl( -2\partial\Lambda(w) \\
&\quad+
\frac{(k+4)(5k+16)}{6(3k+8)(20k^2+93k+102)}\left((12k^2+59k+74)\partial^3 T_\perp(w)
- 48(k+4)\partial T_\perp(w)T_\perp(w)\right)\Bigr).
\end{split}
\end{equation}
Where the fields $T_\perp(w)$ and $\Lambda(w)$ are the following normal ordered polynomials in the generating fields
\begin{equation}\nonumber
\begin{split}
T_\perp(w) &= T(w) + \frac{2}{(3k+8)}J_4(w)J_4(w), \\
(k+2)\Lambda(w) &= -(2k+5)X_4(w)X_4(w)-(2k+5)Y_4(w)Y_4(w)+\frac{4}{(3k+8)}Z_4(w)J_4(w)\\
&\quad+\frac{(k+4)\left(3(k+2)(6k^2+33k+46)\partial^2 T_\perp(w)-(k+4)(11k+26)T_\perp(w)T_\perp(w)\right)}{2(3k+8)(20k^2+93k+102)}\\
&\quad -\frac{8(k+4)}{(3k+8)^2}T_\perp(w)J_4(w)J_4(w)+\frac{8(2k+5)}{3(3k+8)}\partial^2J_4(w)J_4(w)\\
&\quad+\frac{2(2k+5)}{(3k+8)}\partial J_4(w)\partial J_4(w)-\frac{32(2k+5)}{3(3k+8)^3}J_4(w)J_4(w)J_4(w)J_4(w).
\end{split}
\end{equation}
Note that it is a very non-trivial result that OPEs of these fields close with {\em real} coefficients. For example, if we compare the expression for $\Lambda$ with the one of the previous section, we see that contributions of type $\partial^3 J_4(x)$ and $J_4(w)J_4(w)\partial J_4(w)$ disappear. This is essential for closure of our unitary form with real coefficients, and that these terms vanish is a non-trivial computation.

Choosing the adjoint as in the $su(2)$ case, we get no negative norm states for $X_4, Y_4$ and $J_4$ if
$3k+8\geq 0$. Imposing non-negativity of the central charge yields $k+4\leq 15/8$. Hence the possible values for $k$ are
\begin{equation}
 \frac{4}{3}\leq k+4\leq \frac{15}{8}\,.
\label{eq:k}
\end{equation}
Some interesting values of the central charge \eqref{eq:c} in the allowed interval \eqref{eq:k} are $c_4(k=-8/3)=c_4(k=-17/8)=0$ and $c_4(k=-7/3)=c_4(k=-5/2)=1$. The maximal value $c_4=1.105\ldots$ is formally obtained for the irrational value $k=\sqrt{5/2}-4$.

\subsection{The unitary $W^{(2)}_{n}$-algebra}\label{sec:wn}

We finally need to convince ourselves that as far as the operator product algebra of $W^{(2)}_n$ is known, it is consistent with allowing a real unitary form.
Recall the 't Hooft parameter $\lambda=(n-1)(k+n-1)$ \eqref{eq:lambda}.
We then find that
\begin{equation}
\begin{split}
J_n(z)J_n(w) &\sim -\frac{(\lambda-1)}{n(z-w)^2},\qquad
J_n(z)X_n(w)\sim -\frac{Y_n(w)}{(z-w)}, \qquad  J_n(z)Y_n(w)\sim \frac{X_n(w)}{(z-w)}
\end{split}
\end{equation}
and
\begin{equation}
\begin{split}
X_n(z)X_n(w)&\sim Y_n(z)Y_n(w) \sim -\frac{(\lambda-1)}{2(z-w)^n}+
\frac{n(n-1)J_n(w)J_n(w)+2(k+n)T(w)}{2(\lambda-(k+n))(z-w)^{n-2}}\\
&\quad +\frac{n(n-1)\partial J_n(w)J_n(w)+(k+n)\partial T(w)}{2(\lambda-(k+n))(z-w)^{n-3}},\\
X_n(z)Y_n(w) &\sim -\frac{n}{2}\frac{J_n(w)}{(z-w)^{n-1}}-\frac{n}{4}\frac{\partial J_n(w)}{(z-w)^{n-2}}
+\frac{1}{(z-w)^{n-3}}\left( -\frac{n}{12} \partial^2 J_n(w)\right.\\
&\hspace*{-5mm}-\left.\frac{Z_n(w)}{2(\lambda-(k+n))}+\frac{(k+n)n}{2(\lambda-(k+n))\lambda}J_n(w)T_\perp(w)+
\frac{n^3}{12\lambda^2}J_n(w)J_n(w)J_n(w)\right),
\end{split}
\end{equation}
where $T_\perp(w)=T(w)+\frac{n}{2\lambda}J_n(w)J_n(w)$.
We can, as before, search for values of the level with no negative norm states.
The answer in terms of the parameter $\lambda$ is
\begin{equation}
 1\leq \lambda\leq \frac{(2n-1)(n-1)}{n(n-2)}=2+\frac{(n+1)}{n(n-2)}.
\end{equation}
Note that the central charge scales with $n$ for fixed $\lambda$
\begin{equation}
\begin{split}
c &= (1-\lambda)\frac{1-2n+\lambda n \frac{(n-2)}{(n-1)}}{1+\lambda/(n-1)}\sim (\lambda-1)(2-\lambda)\,n\,.
\end{split}
 \end{equation}

\section{Unitary conformal field theories at $\lambda=2$}\label{sec:c1}

Let us consider the case of $k+n=(n+1)/(n-1)$. Then we have $\lambda=2$ and $c=1$.
In this case the algebra compares nicely with the integer lattice CFT rescaled by $\sqrt{n}$, which is strongly generated by  three bosonic fields $h_n(z), e_n(z)$ and $f_n(z)$
of conformal dimension $1, n/2, n/2$.
Their operator products are
\begin{equation}
\begin{split}
h_n(z)h_n(W) &\sim \frac{1}{n} \frac{1}{(z-w)^2},\quad h_n(z)e_n(w)\sim \frac{e_n(w)}{(z-w)},\quad h_n(z)f_n(w)\sim -\frac{f_n(w)}{(z-w)}\,,\\
e_n(z)f_n(w) &\sim \lambda \Bigl( \frac{1}{(z-w)^n}+ \frac{nh_n(w)}{(z-w)^{n-1}}+\frac{\frac{n}{2}\partial h_n(w)+n T(w)}{(z-w)^{n-2}}\\
&\quad+\frac{\frac{n^3}{6}h_n(w)h_n(w)h_n(w)+\frac{n^2}{2}\partial h_n(w)h_n(w)+\frac{n}{6}\partial^2 h_n(w)}{(z-w)^{n-3}}+\dots  \Bigr) \, .
\end{split}
\end{equation}
Here we used an unusual scaling to make the relation to the $W^{(2)}_n$ algebra clear. Indeed this is the operator product algebra of the simple quotient of the level $k+n=(n+1)/(n-1)$ $W^{(2)}_n$ algebra under the identification
\begin{equation}
\Hn = h_n,\qquad \En=e_n\,,\qquad \Fn=f_n\,, \qquad T= \frac{n}{2}h_nh_n\,,\qquad \mathcal W_{n, 3}=0\,.
\end{equation}
For even $n$, these lattice theories describe a free compactified boson, where the compactification radius is $R=n/2$, see for instance \cite{diFrancesco}.

\section{The semi-classical limit and Chern--Simons theory}\label{sec:gravity}

The aim of this section is to derive parts of the relations of the $W^{(2)}_n$-algebra from a gravity point of view.
These computations support the conjecture that the $W^{(2)}_n$-algebra is indeed dual to some higher spin gauge theory
with spin one symmetry.

\subsection{$W^{(2)}_4$ example}

We can conjecturally construct the $W^{(2)}_n$ algebra by quantum Hamiltonian reduction of the
next-to-principal embedding of $sl(2,\mathbb{R})$ into $sl(n,\mathbb{R})$. There is a physical realization of this procedure
at the semi-classical level by studying the asymptotic symmetries of a three dimensional higher spin gauge theory \cite{Henneaux:2010xg,Campoleoni:2010zq,Campoleoni:2011hg}.
The bulk action can be written as the difference of two $sl(n,\mathbb{R})$ Chern--Simons actions,
\bea\label{eq:ch1}
I=\frac{k_{\text{\tiny{CS}}}}{4\pi}\int_\mathcal{M} \left\langle A\wedge dA+\tfrac{2}{3}A\wedge A\wedge A\right\rangle \, .
\eea
The manifold ${\cal M}$ is assumed to be a smooth 3-dimensional manifold with Euclidean signature and cylindrical or
torus boundary. Note however, what we propose is to actually change the contour of the path integral of this Chern--Simons action from the real domain into the complex domain.
It reflects the choice of the real form taken in the boundary algebra,
but importantly it does not change the path integral for the gravitational $sl(2,\mathbb{R})$  part.

We consider the  Chern--Simons bulk theory action in three dimensions with $sl(4,\mathbb{R})$ gauge group.
The Hilbert space of this gravity theory is a representation of the $W^{(2)}_4$ algebra, if we impose certain boundary conditions.
We will see this  in more details below.
There are three non-trivial non-principal embeddings of
$sl(2,\mathbb{R})$ in $sl(4,\mathbb{R})$,
\bea
\underline{15}&\simeq&\underline{3}\oplus\underline{5}\oplus2\,\cdotp\underline{3}\oplus\underline{1}\label{emb1}
\;\simeq\;\underline{3}\oplus3\,\cdotp\underline{3}\oplus3\,\cdotp\underline{1}
\;\simeq\;\underline{3}\oplus4\,\cdotp\underline{2}\oplus4\,\cdotp\underline{1}
\eea
We are  interested here in the first one.
Let $L_a$ be the three $sl(2,\mathbb{R})$ generators with $a=\pm1,0$.
Further let $X_{a}$ and $Y_a$ be six extra spin-2 generators carrying two copies of the three-dimensional representation
of $sl(2,\mathbb{R})$, where $a=\pm1,0$, and denote by $Z_a$ a basis of the five dimensional representations, with $a=\pm2,\pm1,0$,
while $S$ is finally a singlet under $sl(2,\mathbb{R})$.
In this basis, the commutation relations of  $sl(4,\mathbb{R})$ are
\begin{equation}
 \begin{split}
[L_a,L_b] &= (a-b)L_{a+b},\qquad
[S,X_a]= -Y_a,\qquad
[S,Y_a]= X_a,\\
[L_a,X_{b}]   &= (a-b)X_{a+b},\qquad
[L_a,Y_{b}]   = (a-b)Y_{a+b},\qquad
[L_a,Z_b]    =(2a-b)Z_{a+b},\\
[X_{a},X_{b}]&=[Y_{a},Y_{b}]=\tfrac{1}{2}(a-b)L_{a+b},\qquad
[X_{a},Y_{b}]=-Z_{a+b}-4(a^2-\tfrac{1}{3})S\,\delta_{a+b},\\
[Z_a,X_b] &= -\tfrac{1}{6}(a^2+6b^2-3ab-4)Y_{a+b},\qquad
[Z_a,Y_b] = \tfrac{1}{6}(a^2+6b^2-3ab-4)X_{a+b},\\
[Z_a,Z_b] &= \tfrac{1}{12}(a-b)(2a^2+2b^2-ab-8)L_{a+b}\,.
 \end{split}
\end{equation}
We use the notation $\delta_{p+q}:=\delta_{p+q,\,0}=\delta_{p,\,-q}$ for all integer values $p$ and $q$.
Let $\eta$ be the $3\times 3$ matrix with non-zero entries $\eta_{1,-1}=\eta_{-1,1}=1$ and $\eta_{0,0}=-\frac{1}{2}$, and let $\mathcal K$ be the $5\times 5$ matrix with non-zero entries
$\mathcal K_{2,-2}=\mathcal K_{-2, 2}=6$ and $\mathcal K_{1, -1}=\mathcal K_{-1, 1}=-\frac{3}{2}$ and $\mathcal K_{0, 0}=1$. Then the bilinear form $\langle\,,\,\rangle$ is
\begin{equation}
\begin{split}
 \left\langle L_a,L_b\right\rangle&= 2\left\langle X_a,X_b\right\rangle= 2\left\langle Y_a,Y_b\right\rangle=-\eta_{ab},\\
 \left\langle Z_a,Z_b\right\rangle&=-\tfrac{1}{6}\mathcal K_{ab}\qquad\text{and}\qquad
\left\langle S,S\right\rangle=-\tfrac{3}{16}.
\end{split}
\end{equation}
The connection $A$ takes values in the Lie algebra $sl(4,\mathbb{R})$, with the aforementioned $sl(2,\mathbb{R})$ embedding \eqref{emb1} and with the following constant-time boundary condition,
\begin{align}\label{bnc}
A(\varphi)&=g^{-1}a(\varphi)\,g \,d\varphi +g^{-1}\partial_\rho g \,d\rho\,\qquad\text{with}\qquad g=e^{\rho L_0}\qquad\text{and}\nn\\
a(\varphi)&=\hat{k}^{-1}\left(L_1+\mathcal J(\varphi) S+\mathcal L(\varphi) L_{-1}+\mathcal X(\varphi) X_{-1}+\mathcal Y(\varphi) Y_{-1}+\mathcal Z(\varphi) Z_{-2}\right)
\end{align}
where $\mathcal J(\varphi)$, $\mathcal L(\varphi)$, $\mathcal X(\varphi) $, $\mathcal Y(\varphi)$
and $\mathcal Z(\varphi)$ are some arbitrary state dependent functions and $\hat{k}=k_{\text{\tiny{CS}}}/2\pi$.
The boundary conditions \eqref{bnc} are preserved by transformations $A\to A + D\Gamma$ with $\Gamma = g^{-1}\gamma(\varphi)g$ and,
\bea\label{gtfmn}
\gamma(\varphi)=\gamma_S(\varphi)S+\gamma_L^a(\varphi)L_{a}+\gamma_X^a(\varphi)X_{a}+\gamma_Y^a(\varphi)Y_{a}+ \gamma_Z^a(\varphi) Z_{a}\,.
\eea
One can solve all components of $\gamma$ in terms of five free parameters $\gamma_S$, $\gamma_L^1$, $\gamma_X^1$, $\gamma_Y^1$, $\gamma_Z^2$ and the
state dependent functions.
The canonical boundary charges associated with the asymptotic symmetries generated by $\gamma$ are
\begin{align}\label{charge}
Q(\gamma)&=-\frac{k_{\text{\tiny{CS}}}}{2\pi}\int d\varphi\, \left\langle\gamma(\varphi)\,,\,a(\varphi)\right\rangle\\
&=\int d\varphi\left[\tfrac{3}{16}\mathcal J(\varphi)\gamma_S(\varphi)+\mathcal L(\varphi)\gamma^1_L(\varphi)+\tfrac{1}{2}\mathcal X(\varphi)\gamma^1_X(\varphi)+\tfrac{1}{2}\mathcal Y(\varphi)\gamma^1_Y(\varphi)+\mathcal Z(\varphi)\gamma^2_Z(\varphi)\right]\,.\nn
\end{align}
Using the fact that $\{Q(\gamma),a(\varphi)\}=\delta_\gamma a(\varphi)$ and substituting \eqref{charge}  into it, we
find the Poisson brackets between the state dependent functions,
\begin{align}
\nn&\{\mathcal L(\varphi),\mathcal L(\varphi')\} = \mathcal L'\delta-2\mathcal L\delta'-\tfrac{\hat{k}}{2}\delta^{(3)},\qquad
\{\mathcal L(\varphi),\mathcal Z(\varphi')\} =\mathcal Z'\delta-3\mathcal Z\delta',\\
\nn&\{\mathcal L(\varphi),\mathcal X(\varphi')\} = \mathcal X'\delta-2\mathcal X\delta'+\tfrac{1}{\hat{k}}\mathcal J\mathcal Y\delta,\qquad
\{\mathcal L(\varphi),\mathcal Y(\varphi')\} =\mathcal Y'\delta-2\mathcal Y\delta'-\tfrac{1}{\hat{k}}\mathcal J\mathcal X\delta,\\
\nn&\{\mathcal J(\varphi),\mathcal J(\varphi')\}  = -\tfrac{16\hat{k}}{3}\delta',\qquad
\{\mathcal J(\varphi),\mathcal X(\varphi')\} = -\tfrac{16}{3}\mathcal Y\delta,\qquad
\{\mathcal J(\varphi),\mathcal Y(\varphi')\} = \tfrac{16}{3}\mathcal X\delta, \\
\nn&\{\mathcal X(\varphi),\mathcal X(\varphi')\} =\{\mathcal Y(\varphi),\mathcal Y(\varphi')\}=2\mathcal L'\delta-4\mathcal L\delta'+\tfrac{3}{\hat{k}}\mathcal J^2\delta' 
-\tfrac{3}{\hat{k}}\mathcal J\mathcal J'\delta-\hat{k}\delta^{(3)},\\
\nn&\{\mathcal X(\varphi),\mathcal Y(\varphi')\} =-4\mathcal Z\delta-\mathcal J''\delta-3\mathcal J\delta''+3\mathcal J'\delta' 
-\tfrac{4}{\hat{k}}\mathcal J\mathcal L\delta+\tfrac{1}{\hat{k}^2}\mathcal J^3\delta,\\
\nn&\{\mathcal Z(\varphi),\mathcal X(\varphi')\} = -\mathcal Y''\delta
-\tfrac{5}{3}\mathcal Y\delta''+\tfrac{5}{2}\mathcal Y'\delta'+\tfrac{1}{\hat{k}}\mathcal X\mathcal J'\delta+\tfrac{2}{\hat{k}}\mathcal X'\mathcal J\delta-\tfrac{5}{2\hat{k}}\mathcal X\mathcal J\delta'-\tfrac{8}{3\hat{k}}\mathcal Y\mathcal L\delta-\tfrac{1}{\hat{k}^2}\mathcal Y\mathcal J^2\delta,\\
\nn&\{\mathcal Z(\varphi),\mathcal Y(\varphi')\} = \mathcal X''\delta 
+\tfrac{5}{3}\mathcal X\delta''-\tfrac{5}{2}\mathcal X'\delta'+\tfrac{1}{\hat{k}}\mathcal Y\mathcal J'\delta+\tfrac{2}{\hat{k}}\mathcal Y'\mathcal J\delta-\tfrac{5}{2\hat{k}}\mathcal Y\mathcal J\delta'+\tfrac{8}{3\hat{k}}\mathcal X\mathcal L\delta+\tfrac{1}{\hat{k}^2}\mathcal X\mathcal J^2\delta,\\
\nn&\{\mathcal Z(\varphi),\mathcal Z(\varphi')\} = -\tfrac{5}{6}\mathcal L\delta'''+\tfrac{5}{4}\mathcal L'\delta''-\tfrac{3}{4}\mathcal L''\delta'+\tfrac{1}{6}\mathcal L'''\delta
-\tfrac{8}{3\hat{k}}\mathcal L^2\delta'+\tfrac{8}{3\hat{k}}\mathcal L\mathcal L'\delta-\tfrac{1}{\hat{k}}(\mathcal X^2+\mathcal Y^2)'\delta\\&\qquad\qquad\quad\qquad\;+\tfrac{2}{\hat{k}}(\mathcal X^2+\mathcal Y^2)\delta'-\tfrac{\hat{k}}{24}\delta^{(5)}
\end{align}
where $\delta\equiv\delta(\varphi-\varphi')$ and $\delta'\equiv\partial_\varphi\delta(\varphi-\varphi')$ (and similarly for higher derivatives).
We shift,
\bea
\mathcal L\rightarrow \mathcal L+\frac{3}{32\hat k} \mathcal J^2
\eea
such
that all fields are quasi-primaries with respect to $\mathcal L$.
Using the appropriate representation of the delta function and expanding the state dependent functions,
\begin{equation}
\begin{split}
\mathcal L(\varphi) &= -\frac{1}{2\pi}\sum_{p\in\mathbb{Z}} L_p e^{-ip\varphi}\,, \quad\quad \mathcal X(\varphi) = \frac{1}{\pi}\sum_{p\in\mathbb{Z}} X_p e^{-ip\varphi},\\
\mathcal Y(\varphi) &= \frac{1}{\pi}\sum_{p\in\mathbb{Z}} Y_p e^{-ip\varphi}\,,\quad\quad \mathcal J(\varphi) = \frac{i}{2\pi}\sum_{p\in\mathbb{Z}} J_p e^{-ip\varphi},\\
\mathcal Z(\varphi) &= -\frac{i}{2\pi}\sum_{p\in\mathbb{Z}} Z_p e^{-ip\varphi}\,,\quad\quad\delta(\varphi)=\frac{1}{2\pi}\sum_{p\in\mathbb{Z}} e^{-ip\varphi}\,.
\end{split}
\end{equation}
we can rewrite the Poisson brackets in terms of the Fourier modes,
\begin{align}\label{semi-class-algebra}
i\{L_p,L_q\} &= (p-q)L_{p+q} +\tfrac{k_{\text{\tiny{CS}}}}{2}p(p^2-1)\delta_{p+q},\qquad i\{L_p,Z_q\} = (2p-q)Z_{p+q},\nn\\
i\{L_p,X_q\}  &= (p-q)X_{p+q},\qquad
i\{L_p,Y_q\}  = (p-q)Y_{p+q}, \qquad i\{L_p,J_q\} = -qJ_{p+q},\nn\\
i\{J_p,J_q\} &= \tfrac{3k_{\text{\tiny{CS}}}}{16}\,p\,\delta_{p+q} \,,\qquad
i\{J_p,X_q\}  = - Y_{p+q} \,,\qquad
i\{J_p,Y_q\}  =  X_{p+q}\,, \nn\\
i\{X_p,X_q\} &= i\{Y_p,Y_q\}=
\tfrac{1}{2}(p-q)L_{p+q}-\tfrac{12}{k_{\text{\tiny{CS}}}}(p-q)J^2_{p+q}+\tfrac{k_{\text{\tiny{CS}}}}{4}\,p(p^2-1)\delta_{p+q}\nn\\
i\{X_p,Y_q\} &= 
- Z_{p+q}-\tfrac{4}{3}(p^2+q^2-pq-1)J_{p+q} -\tfrac{16}{3k_{\text{\tiny{CS}}}}\left((JL)_{p+q}-\tfrac{88}{9k_{\text{\tiny{CS}}}}J^3_{p+q}\right)\nn\\
i\{Z_p,X_q\}  &= -\tfrac{1}{6}(p^2+6q^2-3pq-4) Y_{p+q}-\tfrac{8}{3k_{\text{\tiny{CS}}}}(LY)_{p+q}\nn\\
&+\tfrac{16}{3k_{\text{\tiny{CS}}}}\left(\tfrac{1}{4}(\partial JX)_{p+q}-\tfrac{3}{4}(\partial XJ)_{p+q}+\tfrac{5}{4}(p-q)(JX)_{p+q}+\tfrac{20}{3k_{\text{\tiny{CS}}}}(J^2Y)_{p+q}\right)\nn\\
i\{Z_p,Y_q\}  &= \tfrac{1}{6}(p^2+6q^2-3pq-4) X_{p+q}+\tfrac{8}{3k_{\text{\tiny{CS}}}}(LX)_{p+q}\nn\\
&+\tfrac{16}{3k_{\text{\tiny{CS}}}}\left(\tfrac{1}{4}(\partial JY)_{p+q}-\tfrac{3}{4}(\partial YJ)_{p+q}+\tfrac{5}{4}(p-q)(JY)_{p+q}-\tfrac{20}{3k_{\text{\tiny{CS}}}}(J^2X)_{p+q}\right)\nn\\
i\{Z_p,Z_q\}  &= \tfrac{1}{12}(p-q)(2p^2-pq+2q^2-8) L_{p+q}+\tfrac{8}{6k_{\text{\tiny{CS}}}}(p-q)L^2_{p+q}-\tfrac{4}{k_{\text{\tiny{CS}}}}(p-q)(X^2+Y^2)_{p+q}\nn\\
&-\tfrac{16}{3k_{\text{\tiny{CS}}}}\left(\tfrac{1}{24}(p-q)(2p^2-pq+2q^2-8) J^2_{p+q}+\tfrac{4}{3k_{\text{\tiny{CS}}}}(p-q)(LJ^2)_{p+q}-\tfrac{16}{9k_{\text{\tiny{CS}}}^2}(p-q)J^4_{p+q}\right)\nn\\
&+\tfrac{k_{\text{\tiny{CS}}}}{24}p(p^2-1)(p^2-4)\delta_{p+q}
\end{align}
where,
\bea
(AB)_p:=\sum_qA_qB_{p-q}=(BA)_p\qquad\text{and}\qquad (\partial AB)_p:=\sum_qqA_qB_{p-q}\,.\nn
\eea
Above, we have shifted  $L_p\rightarrow L_p-\frac{k_{\text{\tiny{CS}}}}{4}\delta_p$ and rescaled $J_p$ properly.
In order to find the algebra at the full quantum level, one should replace $i\{\,,\,\}\rightarrow[\,,\,]$ and introduce normal ordering in nonlinear terms.
We may have a comparison with the OPEs in section 4.2 if  we can identify the bulk $k_{\text{\tiny{CS}}}$ with the boundary level $k$.
Comparing the semiclassical commutator between spin-1 generators and the large $k$ limit of their corresponding OPE suggests that,
\bea
k_{\text{\tiny{CS}}}=-4k \qquad\text{for large }\; k\,.
\eea
The CFT result in section 4.2 should match this result for large $k$.
Specifically the $k_{\text{\tiny{CS}}}$-independent terms in \eqref{semi-class-algebra} should survive after taking this limit in section 4.2.
This suggests the following (large $k$)-rescaling of the spin-3 field in the OPE's,
\bea
Z_4(z)\to Z_4(z)k\,.
\eea
After doing this rescaling we can show that the semi-classical limit of the unitary real form of the $W^{(2)}_4$-algebra in section 4.2  compares nicely to this algebra when $X_p\to2 X_p$ and $Y_p\to2 Y_p$.

\subsection{Operator products from the bulk Chern--Simons theory}

We want to derive parts of the general, classical $\En$, $\Fn$ OPE from the bulk $sl(n)$ Chern--Simons theory, in particular the leading terms.
We use the next-to-principal embedding $n=(n-1)+1$ (eg. from \cite{deBoer:1993iz})
\begin{align}\label{}
    V^{(2)}_{-1}=&-\sum_{l=1}^{n-2}E_{l,l+1},\qquad
    V^{(2)}_{0}=\sum_{l=1}^{n-1}(\frac{n}{2}-l)E_{l,l},\qquad
    V^{(2)}_{1}=\sum_{l=1}^{n-2}l(n-1-l)E_{l+1,l},
\end{align}
where $(E_{ab})_{ij}=\delta_{ai}\delta_{bj}$. The Lie algebra splits into generators $V^{(s)}_m,\ m=-s+1,\ldots,s-1$ with spin $s=2,\ldots,n$ corresponding to the ${\cal W}_{n-1}$ algebra on the CFT side, the $u(1)$ generator $V^{(1)}_0$ and the two spin $n/2$-generators $G^\pm_m$. We take
\begin{align}\label{}
    V^{(1)}_0=&\frac1n\left(\sum_{l=1}^{n-1}E_{l,l}-(n-1)E_{n,n}\right)\ .
\end{align}
Demanding\footnote{As a shorthand, we will sometimes denote $G^\pm_m$ by $V^{(n/2)}_m$ to include it in general formulas like this, even though there are actually three fields of spin $n/2$. We will write them out explicitly when needed.}
\begin{align}\label{}
    [V^{(2)}_m,V^{(s)}_p]=(-p+m(s-1))V^{(s)}_{p+m}
\end{align}
we find
\begin{align}\label{}
    G^+_m=C^+\big(m+\frac{n-2}{2} \big)!\, E_{n/2+m,n},\qquad G^-_m=C^-(-1)^{m+(n-2)/2}\big(m+\frac{n-2}{2} \big)!\, E_{n,n/2-m}.
\end{align}
It is natural to introduce a conjugate and demand
\begin{align}\label{}
\left(G^+_m\right)^\dagger=G^-_{-m}\ ,\qquad \left(V^{(s)}_m\right)^\dagger=V^{(s)}_{-m}\ ,
\end{align}
which will relate $C^+$ and $C^-$. Since the form of the $sl(2)$ algebra is asymmetric, we suggest to use the following non-standard $\mathbb{Z}_2$ operator
\begin{align}\label{}
    \left(A^\dagger\right)_{ij}=(-1)^{i-j}\frac{(i-1)!(n-1-j)!}{(j-1)!(n-1-i)!}A_{ji}
\label{eq:lalapetz}
\end{align}
which is indeed an anti-automorphism of the algebra.
Note here that whenever $i$ or $j$ is equal to $n$ we replace the corresponding factor $(-1)!$ by $-1$ as part of the definition of \eqref{eq:lalapetz}. We could also have chosen to take $(-1)!$ to $1$ since an overall sign change of $G^\pm_m$ is an automorphism.
We now obtain the desired conjugation of $G^\pm_m$ by taking 
\begin{align}\label{}
    C^+=\frac{(C^-)^*}{(n-1)!}\ .
\end{align}
Finally we choose a normalization such that for $s\geq2$
\begin{align}\label{}
    V^{(s)}_{-s+1}=&-\sum_{l=1}^{n-s}E_{l,l+(s-1)}
\end{align}
and we can find $V^{(s)}_{s-1}$ by conjugation
\begin{align}\label{}
    V^{(s)}_{s-1}=&(-1)^{s}\sum_{l=1}^{n-s}\frac{(s+l-2)!(n-1-l)!}{(n-s-l)!(l-1)!}E_{l+(s-1),l} \, .
\end{align}
Indeed, for $V^{(2)}_{\pm1}$ this matches with our embedding.

We now want to derive the boundary OPEs. We will expand around an AdS metric, but only consider one chiral half, so the results should also apply for the Lobachevsky boundary conditions. We will use two methods for the calculation. The first method will only capture the terms proportional to $k_{\text{\tiny{CS}}}$ and the terms independent of $k_{\text{\tiny{CS}}}$, but the calculation is easy, whereas the second method will also capture the non-linear terms, but will be harder.

The first method was used in \cite{Creutzig:2012xb,Creutzig:2013tja} (see also \cite{Ammon:2011ua} for related methods),  and we follow the notation from there. We consider the gauge field $A$ of the Chern--Simons theory as a small deformation of the AdS part
\begin{align}
 A=A_\text{AdS}+\Omega\ ,
\end{align}
where
\begin{align}
 A_\text{AdS} = e^\rho V_1^2 dz + V_0^2 d \rho  ~,
\label{background}
\end{align}
and $\Omega$ solves the linearized equation of motion
\begin{align}\label{eq:eqmomega}
 d\Omega+A_\text{AdS}\wedge\Omega+\Omega\wedge A_\text{AdS}=0\ .
\end{align}
Gauge transformations of $A$ are of the form
\begin{align}
 &\delta A = d \Lambda + [A,\Lambda]~,
\end{align}
and the coupling between the bulk theory operators, $V^{(s)}$, and the boundary fields, which we denote $J^{(s)}$, is defined as
\begin{align}\label{}
    \exp\Big(-\frac{1}{2\pi}\int d^2z [(\Omega_{\bar z})^{(s)}_{s-1}]|_{\textrm{bdry}}J^{(s)}\Big)\ ,
\end{align}
where
\begin{align}\label{}
    \Omega=\sum_{s,m}(\Omega_{\bar z})^{(s)}_{m}V^{(s)}_{m} .
\end{align}
Here $J^{(2)}=T_{{\cal{W}}_{n-1}}$. Also for $G^\pm$ we have chosen a special notation and we denote the dual boundary fields $\En,\Fn$. The coupling in this case then takes the form
\begin{align}\label{}
    \exp\Big(-\frac{1}{2\pi}\int d^2z\,\Big[\frac{\tr([\Omega_{\bar z}]|_{\textrm{bdry}}G^-_{-n/2+1})\En}{\tr(G^+_{n/2-1}G^-_{-n/2+1})}+\frac{\tr([\Omega_{\bar z}]|_{\textrm{bdry}}G^+_{-n/2+1})\Fn}{\tr(G^-_{n/2-1}G^+_{-n/2+1})}\Big]\Big)\, .
\end{align}

Following \cite{Creutzig:2012xb} we claim that the gauge transformation
\begin{align}
\begin{split}\label{eq:gaugetransrepeat}
 \Lambda^{(s)}&=\epsilon_{s}\sum_{n=1}^{2s-1} \frac{1}{(n-1)!} (- \partial)^{n-1}
  \Lambda^{(s)} (z) e^{(s-n)\rho} V^{(s)}_{s-n}
\end{split}
\end{align}
of some operator $\mathcal{O}$ is dual to the transformation
\begin{align}\label{eq:changeonbdry}
    \frac{1}{2\pi i}\oint dz \Lambda^{(s)} (z) J^{(s)\pm}(z)\mathcal{O}(0)
\end{align}
on the CFT side (where the contour encircles $0$), when we only consider linear order of operators, i.e. the part of the OPEs which are independent of $k_{\text{\tiny{CS}}}$. This is simply the analytic continuation of the case where $\Lambda^{(s)}(z)=z^{s-1-m}$, $m=-s+1,\ldots,s-1$, which generate the global transformations.

To calculate the $\En(z)\Fn(0)$ OPE, we first create an insertion $\Fn$ at $z=0$ by using \eqref{eq:gaugetransrepeat} with $\Lambda^{(s)}=1/z$ and $V^{(s)}_{s-n}=G^-_{n/2-1}$ on the AdS solution (the vacuum on the boundary side). Then $\Omega$ takes the following form (here for a general spin field)
\begin{align}\label{eq:omegasol}
    \Omega_z&=\epsilon\frac{1}{(2s-2)!}\del^{2s-1}\Lambda^{(s)}(z)e^{-(s-1)\rho}V^{(s)}_{-(s-1)},\nonumber\\
    \Omega_{\bar z}&=\epsilon\sum_{n=1}^{2s-1} \frac{1}{(n-1)!} (- \partial)^{n-1}
  \bar\del\Lambda^{(s)} (z) e^{(s-n)\rho} V^{(s)}_{s-n}\sim \epsilon 2\pi\delta^{(2)}(z-w)e^{(s-1)\rho}V^{(s)}_{s-1}+\ldots, \\
    \Omega_\rho&=0. \nonumber
\end{align}
We note that the leading term in $\Omega^{(s)}_{\bar z}$ is a delta function, and thus gives the wanted insertion with our bulk/boundary coupling.

We can now find the OPE with $\En$ by performing the transformation \eqref{eq:gaugetransrepeat} with $\Lambda^{(s)}=1,z,\ldots,z^{n-1}$. For $z^{n-1}$ this is not a global symmetry and thus $A_\text{AdS}$ is not invariant. This gives an extra term (after varying both the bulk and the necessary extra boundary term, see \cite{Chang:2011mz})
\begin{align}
 \delta S=-\frac{k_{\text{\tiny CS}}}{2\pi}\int d^2 z e^{2\rho}\tr(\Omega_{\bar z}\delta\Omega_z)
\end{align}
and this gives rise to the central term on the CFT side, proportional to $k_{\text{\tiny{CS}}}$. Note that for the OPE of two spin $s$ generators $A^{(s)}_m$, $B^{(s)}_m$ with dual fields $A^{(s)}$ and $B^{(s)}$, we find that the central term is
\begin{align}\label{}
    A^{(s)}(z)B^{(s)}(0)\sim -(2s-1)k_{\text{\tiny{CS}}}\tr \left(A^{(s)}_{-s+1}B^{(s)}_{s-1}\right)/z^{2s}+\ldots\ ,
\end{align}
which directly shows the need for choosing operators such that the inner product is positive. Notice that the inner product used here simply is the matrix trace. The normalization of the coupling $k_{\text{\tiny{CS}}}$ thus differs from the one used in last subsection.

The procedure can first be done for the generators $V^{(2)}_m$ which we find gives rise to the Virasoro tensor OPE on the boundary side. This fixes the relation between $k_{\text{\tiny{CS}}}$ and the central charge $c-1$ of ${\cal W}_{n-1}$:
\begin{align}\label{}
    k_{\text{\tiny{CS}}}=\frac{c-1}{n(n-1)(n-2)}\ ,
\end{align}
where we have used $\tr \big(V^{(2)}_{-1}V^{(2)}_1\big)=-n(n-1)(n-2)/6$.

Continuing we can examine the OPE of $\Hn= J^{(1)}$ with itself and $\En,\Fn$. Here we find
\begin{align}\label{}
    \Hn(z)\Hn(0)\sim -k_{\text{\tiny{CS}}}\frac{n-1}{n}\frac1{z^2}
\end{align}
\begin{align}\label{}
    \Hn(z)\En(0)\sim\frac{\En}{z}\ ,\qquad\Hn(z)\Fn(0)\sim-\frac{\Fn}{z} \, .
\end{align}
Since $\Hn$ is of spin one, this OPEs will not get $1/k_{\text{\tiny{CS}}}$-corrections. In the classical limit $k\rightarrow\infty$ where $k_{\text{\tiny{CS}}}\simeq-k$ this fits with the bulk side.

For the $\En(z)\Fn(0)$ OPE we find that for $\Lambda^{(n/2)} (z)=z^{p-1}$ with  $p=1,\ldots,n-1$ we have
\begin{align}\label{}
    \delta_{\Lambda}\Omega_{\bar z}=&(-1)^{s+1}|C^-|^2\frac{(p-1)!}{(n-1)!}\sum_{s=1}^{n-p}\frac{1}{(n-s-p)!}\sum_{t=n-p-(s-1)}^{n-1-(s-1)}\frac{(t+s-2)!(n-t-1)!}{(t-n+s-1+p)!(n-t-s)!}\nonumber\\
    &\qquad\qquad\times\left(E_{t+(s-1),t}-\delta_{s,1}E_{n,n}\right)e^{(s-1)\rho}(-\partial)^{n-s-p}2\pi\delta^{(2)}(z)\nonumber\\
    \sim &-|C^-|^2\frac{(p-1)!}{(n-1)!}V^{(s)}_{s-1}e^{(n-p-1)\rho}2\pi\delta^{(2)}(z)+\cdots\ ,
\end{align}
where we have displayed the highest spin term explicitly. In order to get this result one has to remember to keep only terms that are not vanishing at the boundary and we have already made sure to only keep terms that have no $z^a\delta^{(2)}(z)$ behaviour with $a$ positive after partial differentiation in the bulk-boundary coupling.

It is hard to derive the non-leading spin terms, so we simply show the result for the leading spin
\begin{align}\label{}
    \En(z)\Fn(0)\sim\frac{-k_{\text{\tiny{CS}}}|C^-|^2}{z^n}-\frac{|C^-|^2\tfrac{n}{n-1}\Hn}{z^{n-1}}-\sum_{p=1}^{n-2}\frac{\tfrac{(p-1)!}{(n-1)!}|C^-|^2J^{(n-p)}+\cdots}{z^{p}}\ .
\end{align}
This fixes the normalization $C^-$ in the classical limit
\begin{align}\label{}
    |C^-|^2=\frac{\lambda_{n-1}}{k}\ .
\end{align}
Had we chosen the opposite sign on the conjugation of $G^+_m$ in \eqref{eq:lalapetz}, then we would have a minus sign on the right hand side of this equation. However this is the natural choice for the classical limit where $\lambda_{n-1}/k\sim(n-1)!k^{n-2}$ is positive. We also note that this normalization is unnatural from the bulk perspective since it contains the level. It seems more sensible to take a simple normalization like e.g. $|C^-|^2=n-1$ (which is always positive). When we then define $X^{(\text{\tiny bulk})}_m=i(G^+_m+G^-_m)/2$ and $Y^{(\text{\tiny bulk})}_m=(G^+_m-G^-_m)/2$, we have the simple relation that $X^{(\text{\tiny bulk})}_m$ and $Y^{(\text{\tiny bulk})}_m$ are the generators dual to the boundary fields $X_n$ and $Y_n$. This shows that the normalization in \eqref{eq:scarlett} is quite natural from the bulk viewpoint.

Let us now discuss how to obtain the normal ordered terms in the OPE. We use the method from \cite{Gutperle:2011kf,Moradi:2012xd,Creutzig:2013tja}. Let $a$ be the gauge field with the $\rho$-dependence adjointly removed
\begin{align}\label{}
  A=b^{-1}ab+b^{-1}db\ ,\qquad b=e^{\rho V^{(2)}_0}\ .
\end{align}
As before, we first turn on a background with some current $-\int d^2 z\mu(z,\bar z) J^{(t)}$. Here $\mu=\epsilon 2\pi\delta^{(2)}(z-w)$ compared to \eqref{eq:omegasol}. With this gauge field turned on, the the lowest weight gauge takes the following form
\begin{align}\label{}
  a_z=& V^{(2)+}_1-\frac{1}{k_{\text{\tiny{CS}}}}\left(\langle J^{(s)}\rangle_0+\langle J^{(s)}\rangle_\mu\right) g^{V^{(s')}_{-s'+1}\,V^{(s)}_{s-1}}V^{(s')}_{-s'+1}    \ , & \nonumber \\
  a_{\bar z}=& \sum_{m=-t+1}^{t-1} \mu_m V^{(t)}_m \ .
\end{align}
Here $\langle J^{(s)}\rangle_0$ is independent of $\mu$ and thus holomorphic, and $\langle J^{(s)}\rangle_\mu$ is first order in $\mu$. The inverse of the trace metric is denoted $g^{V^{(s')}_{-s'+1}\,V^{(s)}_{s-1}}$. Finally $\mu\equiv\mu_{t-1}$. The idea is now to solve the equations of motion
\begin{align}\label{}
\del a_{\bar z}-\bar\del a_z+[a_z,a_{\bar z}]=0
\end{align}
to linear order in $\mu$ for $\bar\del \langle J^{(s)}\rangle_\mu$. This is then equivalent to taking the anti-holomorphic partial derivative of the OPE on the boundary side:
\begin{align}\label{}
-\frac{1}{2\pi}\del_{\bar z}\int\, d^2 w\mu(w,\bar w)\langle J^{(s)}(z)J^{(t)}(w)\rangle\ .
\end{align}
We can now use this for the $\En(z)\Fn(0)$ OPE. In general this is a hard problem, however, if we focus on the terms that only depend on the spin one field, we get the simpler equation
\begin{align}\label{}
    \bar\del \langle \En\rangle_{\mu}=-\frac{k_{\text{\tiny{CS}}}}{(n-2)!}\tr\left(G^{-}_{-(n-2)/2}G^{+}_{(n-2)/2}\right)\left(\del+\tfrac{1}{k_{\text{\tiny{CS}}}}\langle J^{(0)}\rangle g^{V^{(1)}_{0}\,V^{(1)}_{0}}\right)^{n-1}\mu \, .
\end{align}
Using this we obtain the most singular terms in the OPE
\begin{align}\label{}
    \En(z)\Fn(0)\sim&-\frac{k_{\text{\tiny{CS}}}|C^-|^2}{z^n}-\frac{|C^-|^2\tfrac{n}{n-1}\Hn}{z^{n-1}}\nonumber\\
    &+|C^-|^2\frac{-\tfrac{1}{(n-1)(n-2)}T_{{\cal W}_{n-1}}+\tfrac{n}{2(n-1)}\del\Hn-\tfrac{1}{k_{\text{\tiny{CS}}}}\tfrac{n^2}{2(n-1)^2}\Hn^2}{z^{n-2}}+\cdots\ .
\end{align}
If we remember that the total Virasoro tensor is
\begin{align}\label{}
    T=T_{{\cal W}_{n-1}}-\frac{n}{2(n-1)k_{\text{\tiny{CS}}}}\Hn^2\
\end{align}
this matches perfectly with the first three terms in the classical limit of \eqref{eq:angelinajolie}.

\section{Conclusion and Outlook}

The main result of this work is that we conjectured that the $W^{(2)}_n$ algebra for even $n$ allows a unitary real form, and that for a certain range of levels it even can have a positive definite inner product. We verified that the conjecture is consistent with the known operator product algebra. For the special value $\lambda=2$ of the 't~Hooft parameter, we then indeed found a unitary lattice CFT corresponding to the simple quotient of the $W^{(2)}_n$ algebra at level $k+n=(n+1)(n-1)$ (for even $n$).

We then proposed an 't~Hooft limit that is very much analogous to the original minimal model holography \cite{Gaberdiel:2012uj}, and we verified for some leading terms that Chern--Simons theory on $SL(n;\RR)$ with next to principal boundary conditions indeed reproduces the known operator product.

As mentioned in the introduction, there is a zoo of quantum Hamiltonian reductions, and there are also many more which only have one current of dimension one. It is an obvious question whether these algebras also allow a unitary real form, whether they also allow for certain levels with a positive inner product and whether some of these levels even allow for a unitary CFT. However, since almost nothing is known about these algebras it will not be easy to address these questions. A more realistic problem is to study a supersymmetric analog. In \cite{Creutzig:2011np, Creutzig:2011cu, Alfes:2012pa} superalgebras containing the $W^{(2)}_n$-algebra have been found and studied. Supersymmetric algebras containing the algebra of Feigin and Semikhatov
with $\mathcal N=2$ superconformal structure appear as the quantum Hamiltonian reductions of the affine Lie super algebra of $psl(n|n)$ for a next to principal embedding of $sl(2)$. By this, we mean an embedding where $sl(2)$ is principal in one of the two $sl(n)$ subalgebras and next to principal in the other one.
Such reductions fall into the framework of  \cite{MR2060475}. The resulting algebra is expected to contain the extended $\mathcal N=2$ superconformal algebra of Kazama-Suzuki cosets \cite{Kazama:1988qp} as subalgebra. Thus, this algebra might add another supersymmetric higher spin/CFT holography to the existing ones \cite{Creutzig:2011fe, Candu:2012jq, Creutzig:2012ar, Beccaria:2013wqa, Gaberdiel:2013vva, Creutzig:2013tja}. We consider it to be probable that this supersymmetric version behaves even better from the unitarity point of view than the algebra presented in this work.

In this paper, we have focused on the symmetry algebra and its unitary representations.
In order to construct a unitary theory, we may need to set singular vectors to zero. However, typically these singular vectors cannot be seen from the classical limit of the gravity theory. Therefore, the analysis on the unitarity should give some insights on quantum effects of higher spin gravity theory, and it is worthwhile to investigate this further from the gravity viewpoint.
For the analysis of symmetry algebra, it does not matter whether the gravity theory includes massive matter fields or not since they are irrelevant near the boundary. It is indeed a nice point that higher spin gravity can be defined only by Chern--Simons gauge theories. However, for the known case by \cite{Gaberdiel:2010pz}, the gravity theory includes massive scalars along with the Chern--Simons gauge fields, and the existence of the massive scalars is essential for the duality to hold. This implies that massive matter fields could also be important even in the present case.

\section*{Acknowledgments}

HA and DG thank Michael Gary, Radoslav Rashkov, Max Riegler and Jan Rosseel for discussions.
HA was supported by the Dutch stichting voor Fundamenteel Onderzoek der Materie (FOM).
TC was supported by a start-up research grant of the University of Alberta.
DG was supported by the START project Y~435-N16 and projects I~952-N16 and I~1030-N27 of the Austrian Science Fund (FWF).
YH was supported in part by JSPS KAKENHI Grant Number 24740170.
PR was supported by AFR grant 3971664 from Fonds National de la
Recherche, Luxembourg, and partial support by the Internal Research Project GEOMQ11
(Martin Schlichenmaier), University of Luxembourg, is also acknowledged.


\providecommand{\href}[2]{#2}\begingroup\raggedright\endgroup

\end{document}